\documentclass[twocolumn,noshowpacs,preprintnumbers,amsmath,amssymb]{revtex4}
\usepackage{graphicx}% Include figure files
\usepackage{dcolumn}% Align table columns on decimal point
\usepackage{bm}% bold math

\begin{document}
%\preprint{APS/123-QED}
%\baselineskip 24pt
\title{Constant-temperature molecular-dynamics  algorithms for mixed hard-core/continuous potentials}
\author{Yao A. Houndonougbo and Brian B. Laird\footnote{Author to whom correspondence should be addressed}}
%Department of Chemistry\\ University of Kansas \\ Lawrence, KS 66045, USA}
%\author{Yao A. Houndonougbo and Brian B. Laird}
\affiliation{Department of Chemistry\\ University of Kansas \\ Lawrence, KS 66045, USA}
\date{\today}
\begin{abstract}
We present a set of second-order, time-reversible algorithms for the isothermal ($NVT$) 
molecular-dynamics (MD) simulation of systems with mixed hard-core/continuous potentials. The 
methods are generated by combining
real-time Nos\'e thermostats with our previously developed Collision Verlet algorithm 
[Mol. Phys. {\bf 98}, 309 (1999)] for constant energy MD simulation of such systems. 
In all we present 5 methods, one based on the Nos\'e-Hoover [Phys. Rev. A {\bf 31}, 1695 (1985)] 
equations of motion and four based on the Nos\'e-Poincar\'e [J.Comp.Phys., {\bf 151} 114 (1999)] 
real-time formulation of Nos\'e dynamics.  The methods are tested using a system of hard spheres 
with attractive tails and all correctly reproduce a canonical distribution of instantaneous 
temperature.  The Nos\'e-Hoover based method and two of the Nos\'e-Poincar\'e methods are shown to have good energy conservation in long simulations. 
\end{abstract}
\maketitle

\section{Introduction}
Algorithms for molecular-dynamics simulation can be generally be divided into
two distinct classes depending upon the nature of the potential\cite{Allen87}. 
For systems with 
continuously differentiable potentials, the trajectory is generated through
the numerical integration of the equations of motion - a coupled set of
differential equations - typically with a fixed time step. At the other end 
of the spectrum are methods for discontinuous potentials, such as hard spheres or
the square-well potential.  Such algorithms are event driven in that the system
is advanced ballistically between "collisions", which are then resolved exactly. 
There exist, however, model interaction potentials of theoretical and practical
importance that are hybrids of continuous and discontinuous potentials - for 
example, the restricted primitive model for electrolyte solutions or the dipolar 
hard-sphere model of polar fluids. To date, simulation studies for such systems
have primarily been restricted to Monte Carlo studies due to the lack of
a viable molecular-dynamics (MD) algorithm. To remedy this, we have recently
introduced a new molecular-dynamics method for such systems~\cite{Houndonougbo00}.  
The algorithm, referred to as Collision Verlet,  
has good energy conservation and is far more stable over long time
simulation than previous integrators for hybrid continuous/discontinous 
systems. The Collision Verlet algorithm was formulated as a constant
energy simulation method, which generates configurations from a microcanonical
($NVE$) distribution. However, to mimic experimental conditions most modern
simulations are run under isothermal ($NVT$) or isothermal/isobaric ($NPT$)
conditions. In this work, we introduce and evaluate several reformulations
of Collision Verlet to generate trajectories whose phase space points
are canonically (isothermally) distributed.

The NVT (isothermal) Collision Verlet algorithms developed here are all based on 
the extended Hamilitonian of Nos\'{e}\cite{Nose84a}, which is a standard technique
for generating canonical trajectories for the simulation of systems with continuous 
interaction potentials.
In the Nos\'e approach, the phase space of the system is augmented by
the introduction of an auxilliary variable $s$ and its conjugate momentum $\pi$ 
(with ``mass'' $Q$). For a system with  a potential $V$, the Nos\'{e} extended 
Hamiltonian is
\begin{equation}
{\cal H}_{nos \acute{e}} = \sum_{i} \frac{\tilde{p}_{i}^2}{2m_{i}s^2}+
V({\bf q}) + \frac{\pi^2}{2Q} + gkT\ln s,
\label{HNose}
\end{equation}
where $\tilde{p}_i$ is the momentum conjugate to the position $q_i$ and is related
to the actual momentum , $p_i$, by the relation $p_i = \tilde{p}_i/s$, and 
the parameter $g = N_f + 1$, where $N_f$ is the number of 
degrees of freedom of the system. With this choice of $g$, it can be readily shown\cite{Nose84a},
assuming ergodicity, that constant energy (microcanonical) dynamics generated
by the Nos\'e Hamiltonian produces a canonical (constant temperature) distribution in the 
reduced phase space $\{{\bf \tilde{p}}/s,{\bf q}\}$.  

The generation of phase space configurations distributed in the canonical 
ensemble within the Nos\'e dynamical scheme is accomplished by a dynamical
rescaling of time. The real time of the simulation, $t$, is related to the
Nos\'e time, $\tau$, by the transformation $\frac{d \tau}{d t} = s$. Since
numerical integration methods  generally operate with a fixed time step,
the transformation to real time generates a nonuniform grid of time points\cite{Nose84b},
which is inconvenient for the calculation of system averages. To remedy this,
two schemes have been developed to produce equations of motion for 
Nos\'e dynamics that generate trajectories directly in real time. 
By applying time and coordinate transformations directly to the Nos\'e equations
of motion Hoover\cite{Hoover85}, derived a set of real-time equations of motion
for Nos\'e dynamics, defining  the so-called Nos\'{e}-Hoover method. This approach
has become the most widely isothermal simulation method,
but has a drawback in that the coordinate transformation used is not
canonical and the Nos\'e-Hoover equations of motion are non-Hamiltonian in structure,
precluding the use of symplectic integration schemes\cite{Sanz-Serna95}. In an alternate
approach Bond, Leimkuhler and Laird\cite{Bond99} apply a Poincar\'e time transformation
to the Nos\'e Hamiltonian to give the so-called Nos\'e-Poincare Hamiltonian, from
which real-time, fully Hamiltonian equations of motion for Nos\'e dynamics are
generated. 

In this work we present constant temperature simulation methods for mixed continuous/discontinuous
interaction potentials generated by adapting the Collision Verlet method within  both
the Nos\'e-Hoover and Nos\'e-Poincar\'e schemes. In the next section we briefly review the 
standard Collision Verlet algorithm\cite{Houndonougbo00} followed by 
the introduction of the Nos\'e-Hoover Collision Verlet (NHCV) and Nos\'e-Poincar\'e Collision 
Verlet (NPCV) algorithms in Sections 3 and 4, respectively. The algorithms are evaluated
in Section 5 through numerical experiments on a model system. In section 6, we conclude.

\section{The Collision Verlet Algorithm}

In this section we review the Collision Verlet\cite{Houndonougbo00} algorithm for the numerical
integration of the dynamics of systems with mixed continuous/discontinuous interaction potentials.
We consider $N$ particles interacting through a continous potential plus a hard core, assumed 
here to
be spherical. To facilitate the construction of numerical methods, it is useful to describe the
dynamics of the system within a Hamiltonian format, but for a system with a discontinuous potential
the construction of a Hamiltonian as the generator of the dynamical equations of motion is 
problematic. In this work, we observe that the hard sphere interaction potential, $V_{hs}(\{ {\bf q}\})$
typically can be approximated to any degree of accuracy by a sequence of steeply repulsive
continuous functions. In this sense, the energy function ${\cal H}$ of the mixed system
is refered to here as a pseudo-Hamiltonian. Here the pseudo-Hamiltonian is given by
\begin{equation}
{\cal H} = T({\bf p})+ V_{hs}(\{{\bf q}\}) + V_c(\{{\bf q}\}),
\end{equation}
where the kinetic energy $T({\bf p}) = \sum_i \frac{p_{i}^2}{2m_{i}}$,
$V_{hs}(\{{\bf q}\})$ is the hard sphere potential and $V_c(\{{\bf q}\})$ is a 
continuously differentiable potential energy function, that we assume to be pairwise
additive, that is,
\[
V_c(\{{\bf q}\}) = \sum_{i} \sum_{j>i}  v_c(q_{ij}) \; ,
\]
where $v_c$ is a pair potiential, $q_{ij}$ is the distance between two particles
indexed by $i$ and $j$, and the sum is over all pairs of particles.

The Collision Verlet algorithm is based on the splitting of the continuous
pair potential, $v_c(q)$, into a short range part, $v_1(q)$, and a long range 
part, $v_2(q)$, according to
\begin{equation}
v_c(q) = v_1(q) + v_2(q)
\end{equation}
The potential splitting is rendered so that the force due to the long-range part of the
potential vanishes at the hard-sphere contact distance (i.e. $v^{\prime}_2(\sigma) = 0$).
This form of the potential splitting is necessary for the construction of a second-order method - 
For the motivation and specific details of this splitting technique the reader is
referred to reference\cite{Houndonougbo00}. 
The pseudo-Hamiltonian is then split acoordingly. For 
generality, let consider ${\cal H}$ as a pseudo-Hamiltonian of any given mixed 
impulsive-continuous system. Next, we partition ${\cal H}$ in the following way:
\begin{equation}
 {\cal H} = {\cal H}_1 + {\cal H}_2\; ,
\label{generalSplit}
\end{equation}
where ${\cal H}_1$ includes the kinetic energy, the hard sphere potential, $V_{hs}$, 
and the short range potential, $V_1$; ${\cal H}_2$ must include the long
range potential, $V_2$. 
A Trotter factorization\cite{Sanz-Serna95} then gives the following approximation for the
dynamical flow map, $\phi_{\cal{H}}(\tau)$, defined as the operator (associated with the
Hamiltonian ${\cal H}$) that advances the phase
space configuration a time $\tau$ into the future,
\begin{equation}
\phi_{\cal{H}}(\tau) = \phi_{{\cal H}_2}(\frac{\tau}{2})\phi_{{\cal H}_1}(\tau)\phi_{{\cal H}_2}
(\frac{\tau}{2})
\label{trotter}
\end {equation}
Since ${\cal H}_2$ only contains the long-range potential, the flow map $\phi_{{\cal H}_2}$ can be
constructed exactly.  The flow map corresponding to ${\cal H}_1$ is approximated in the following 
way
\begin{equation}
\phi_{{\cal H}_1}  \approx \phi_{\small T + V_1}(\tau_c^{n_c+1}) \prod_{i=1}^{n_c} 
[ \phi_{\small V_{hs}} \phi_{\small T+ V_1}(\tau_c^{n_c+1 - i})] \;
\label{H1flowapprox}
\end {equation}
where $n_c$ is the number of hard-sphere collisions during the time
step $h$, $\tau_i^{(c)}$ is the time between each collision (with $\tau_1^{(c)}$ being
measured from the beginning of the time step until the first collision
and $\tau_{n_c+1}^{(c)}$ measured from the last collision to the end of the time
step so that $\sum_{i=1}^{n_c+1} = \tau$), and $\phi_{V_{hs}}$ is the flow map for an instantaneous 
hard-sphere collision. Finally, the flow map for the motion of the particle between collisions 
is approximated using the St\"omer-Verlet algorithm generated by a further Trotter factorization
\begin{equation}
\phi_{\small T+ V_1}(\tau)  \approx \phi_{\small V_1}(\frac{\tau}{2})  \phi_{\small T}(\tau) \phi_{\small V_1}(\frac{\tau}{2}) \;.
\end{equation}

The most CPU intensive part of the Collision Verlet algorithm is the determination of
the time to next collision $\tau_c$.  The collision condition for two particles 
$i$ and $j$ can be written as
\begin{equation}
\|\mbox{\bm$q$}_i(\tau_c) - \mbox{\bm$q$}_j(\tau_c)\|^2 -
\sigma^2 = 0\;. \label{coll_cond} 
\end{equation}
Since the trajectories between collisions are approximated within the Collision 
Verlet scheme by quadratic equations, the collision condition (\ref{coll_cond}) is
a quartic equation.  To ensure that all collisions are resolved correctly, it is necessary to 
accurately resolve the smallest positive root to this quartic equation.
This is not a trivial problem as the root becomes increasingly unstable as smaller time steps 
are used (i.e., when the time to collision is small). To increase efficiency and accuracy of the
computation, we employed in all the 
simulations in this paper a root finding method based on Cauchy indices\cite{Henrici74}. The details 
of the collision-time calculation are given in the Appendix.

\section{Collision Verlet with a Nos\'{e}-Hoover thermostat}

As discussed in the introduction, the Nos\'e-Hoover method for isothermal molecular-dynamics 
simulation is generated by applying time and coordinate transformations to the equations of 
motion generated by the Nos\'e Hamiltonian (Eq.~\ref{HNose}), which are
\begin{equation}
\frac{d{q}_i}{d\tau} = \frac{\tilde{p}_{i}}{m_{i}s^2},\; \;  \frac{d s}{d \tau} = \frac{\pi}{Q},
\label{eqMotNose}
\end{equation}
\begin{equation}
\frac{d\tilde{p}_i}{d\tau} =-\frac{\partial}{\partial q_{i}}V_{c}(q), \; \;
\frac{d \pi}{d \tau} = \sum_{i}\frac{\tilde{p}_{i}^2}{m_{i}s^3} - \frac{gkT}{s}  \;.
\label{eqMotNosesecond}
\end{equation}
Conversion to real time, $t$, is accomplished through the following transformations
\begin{equation}
{\bf p} = \frac{\tilde{\bf p}}{s}, \; \; 
\frac{d \tau}{d t}  = {s} .
\label{noseTransf}
\end{equation}
In addition, Hoover simplified the resulting equations of motion 
by introducing a further variable tranformation
\begin{equation}
\eta =\ln s\; \; \xi = \dot{\eta}
\end{equation}
resulting in the so-called Nos\'{e}-Hoover equations of motion:
\begin{equation}
\dot{q}_{i} = \frac{p_{i}}{m_{i}},\; \; \dot{p}_{i} =-\frac{\partial}{\partial q_{i}}V(q)-p_{i}\xi,
\label{conteqMotNHfirst}
\end{equation}
\begin{equation}
\dot{\eta} = \xi, \; \; \dot{\xi} = \frac{1}{Q}\left(\sum_{i}
\frac{p_{i}^2}{m_{i}} - gkT\right) \; .
\label{conteqMotNH}
\end{equation}
These equations of motion can be shown to generate configurations distributed according to an isothermal
(canonical) distribution as long as the system is ergodic and $g = N_f$, the number of degrees
of freedom.\\
Since the coordinate transformation is non-canonical, the equations of motion are not
derivable from a Hamiltonian, however a conserved energy does exist and is given by 
\begin{equation}
E =  \sum_{i} \frac{p_{i}^2}{2m_{i}}+V(q)+\frac{1}{2} Q \xi^2 + gkT\eta.
\end{equation}

In order to simplify the construction of splitting methods for this non-Hamiltonian system and
to make contact with the earlier literature, we write the flow map in terms of 
a Liouville operator, $\cal{L}$, as follows
\begin{equation}
\phi(\tau) = e^{\cal{L}} \;.
\end{equation}
The Liouville operator corresponding to the Nos\'{e}-Hoover equations of motion above
is
\begin{eqnarray}
{\cal L}& =& \sum_{i}\frac{p_{i}}{m_{i}}\frac{\partial}{\partial q_{i}} + {\cal L}_{hs}
- \sum_{i}p_{i}\xi \frac{\partial}{\partial p_{i}}
-\sum_{i}\frac{\partial}{\partial q_{i}}V(q)
\frac{\partial}{\partial p_{i}} \nonumber \\
&&+ \xi\frac{\partial}{\partial \eta}
+\frac{1}{Q}\left(\sum_{i}\frac{p_{i}^2}{m_{i}} - gkT\right)
\frac{\partial}{\partial \xi}\;,
\label{liouOp}
\end{eqnarray}
where we have explicitly included a hard-sphere term, ${\cal L}_{hs}$

To get a reversible method for the Nos\'e-Hoover method with mixed potentials, the above Liouville 
operator is split in the following way:
\begin{equation}
{\cal L} = {\cal L}_{1} + {\cal L}_{2} + {\cal L}_{3},
\label{liouOpSplit}
\end{equation}
with
\begin{equation}
{\cal L}_{1} = {\cal L}_{hs} + \sum_{i}\frac{p_{i}}{m_{i}}\frac{\partial}{\partial q_{i}}
-\sum_{i}\frac{\partial}{\partial q_{i}}V_1(q)
\frac{\partial}{\partial p_{i}},
\label{liouOpiLfirst}
\end{equation}
\begin{equation}
{\cal L}_{2} = -\frac{\partial}{\partial q_{i}}V_2(q)\frac{\partial}{\partial p_{i}}
\label{liouOpiLsnd}
\end{equation}
and
\begin{equation}
{\cal L}_{3} = - \sum_{i}p_{i}\xi\frac{\partial}{\partial p_{i}}
+\frac{1}{Q}\left(\sum_{i}\frac{p_{i}^2}{m_{i}} - gkT\right)
\frac{\partial}{\partial \xi} + \xi\frac{\partial}{\partial \eta}.
\label{liouOpiL3}
\end{equation}

%\begin{equation}
%{\cal L} = {\cal L}_{1} + {\cal L}_{2} ,
%\label{liouOpSplitNH}
%\end{equation}
%with
%\begin{equation}
%{\cal L}_{1} = \sum_{i}\frac{p_{i}}{m_{i}}\frac{\partial}{\partial q_{i}} + {\cal L}_{hs}
%-\sum_{i}\frac{\partial}{\partial q_{i}}V_1(q)
%\frac{\partial}{\partial p_{i}},
%\label{liouOpiLfirstNH}
%\end{equation}
%and
%\begin{equation}
%{\cal L}_{2} = - \sum_{i}p_{i}\xi \frac{\partial}{\partial p_{i}}
%-\frac{\partial}{\partial q_{i}}V_2(q)\frac{\partial}{\partial p_{i}}+\frac{1}{Q}\left(\sum_{i}\frac{p_{i}^2}{m_{i}} - gkT\right)
%\frac{\partial}{\partial \xi} + \xi\frac{\partial}{\partial \eta}.
%\label{liouOpiL3NH}
%\end{equation}
A Trotter factorization is now applied to this splitting. 
%\begin{equation}
%e^{{\cal L}\tau} \approx  e^{{\cal L}_2\tau/2} e^{{\cal L}_1\tau} e^{{\cal L}_2\tau/2} 
%\end{equation}
\begin{equation}
e^{\displaystyle {\cal L}\tau} = e^{\displaystyle {\cal L}_3\tau/2}e^{\displaystyle {\cal L}_2\tau/2}
e^{\displaystyle {\cal L}_1\tau}e^{\displaystyle {\cal L}_2\tau/2}e^{\displaystyle {\cal L}_3\tau/2} + {\cal O}(\tau^3)\; .
\label{trotterNH}
\end {equation}
The operator $e^{{\cal L}_1\tau}$  is approximated using the Collision Verlet
method described in the previous section - see Eq.~\ref{H1flowapprox}. The solution of the 
operator $e^{{\cal L}_2\tau/2}$ is straightforward. To
find the solution of the operator $e^{{\cal L}_3\tau/2}$,i.e,
\begin{equation}
\left(\begin{array}{c}q_{i,n+1} \\ p_{i,n+1} \\ \eta_{n+1} \\ \xi_{n+1}\end{array} \right) = 
e^{{\cal L}_3\tau/2} \left(\begin{array}{c}q_{i,n} \\ p_{i,n} \\ \eta_n \\ \xi_{n} \end{array}\right)
,
\end{equation}
%we use an explicit time-reversible integration\cite{Holian90,Jang97}
%based on the St\"{o}mer-Verlet method, giving the following equations
we further split ${\cal L}_3$. That is,
\begin{equation}
{\cal L}_3 = {\cal L}_3^{(1)} + {\cal L}_3^{(2)},
\label{L3Split}
\end{equation}
with
\begin{equation}
{\cal L}_3^{(1)} =- \sum_{i}p_{i}\xi\frac{\partial}{\partial p_{i}} 
+ \xi\frac{\partial}{\partial \eta},
\end{equation}
and 
\begin{equation}
{\cal L}_3^{(2)}=\frac{1}{Q}\left(\sum_{i}\frac{p_{i}^2}{m_{i}} - gkT\right)
\frac{\partial}{\partial \xi}.
\end{equation}
The corresponding Trotter factorization of this splitting is
\begin{equation}
e^{{\cal L}_3\tau} \approx  e^{{\cal L}_3^{(2)}\tau/2} e^{{\cal L}_3^{(1)}\tau}
 e^{{\cal L}_3^{(2)}\tau/2}.
\end{equation}
The solution of the operator $e^{{\cal L}_3^{(2)}\tau/2}$ is straightforward. 
The operator $e^{{\cal L}_3^{(1)}\tau}$ is solve from a further splitting.
 The solution of the operator $e^{{\cal L}_3\tau/2}$ gives
\begin{equation}
\xi_{n+1/2} = \xi_{n} + \frac{\tau}{4Q}\left(\sum_i
\frac{(p_{i,n})^2}{m_{i}} - gkT\right),
\end{equation}
\begin{equation}
\eta_{n+1} = \eta_{n} + \frac{\tau}{2}\xi_{n+1/2},
\end {equation}
\begin{equation}
p_{i,n+1} =  p_{i,n}\frac{1-\tau\xi_{n+1/2}/4}{1+\tau\xi_{n+1/2}/4},
\end{equation}
\begin{equation}
\xi_{n+1} = \xi_{n+1/2} + \frac{\tau}{4Q}\left(\sum_i
\frac{(p_{i,n+1})^2}{m_{i}} - gkT\right).
\end{equation}
The algorithm is tested in Section 5 
for a system of hard spheres with inverse-sixth-power attractive tails.

Certainly, the Liouville operator splitting used above is not the only
possible method. For example, another splitting is 
\begin{equation}
{\cal L} = {\cal L}_{1} + {\cal L}_{2} ,
\label{liouOpSplitNH}
\end{equation}
with
\begin{equation}
{\cal L}_{1} = \sum_{i}\frac{p_{i}}{m_{i}}\frac{\partial}{\partial q_{i}} + {\cal L}_{hs}
-\sum_{i}\frac{\partial}{\partial q_{i}}V_1(q)
\frac{\partial}{\partial p_{i}},
\label{liouOpiLfirstNH}
\end{equation}
and
\begin{eqnarray}
{\cal L}_{2} = - \sum_{i}p_{i}\xi \frac{\partial}{\partial p_{i}}
-\frac{\partial}{\partial q_{i}}V_2(q)\frac{\partial}{\partial p_{i}}\nonumber\\
+\frac{1}{Q}\left(\sum_{i}\frac{p_{i}^2}{m_{i}} - gkT\right)
\frac{\partial}{\partial \xi} + \xi\frac{\partial}{\partial \eta}.
\label{liouOpiL3NH}
\end{eqnarray}
can be used. Using a Trotter factorization gives
\begin{equation}
e^{{\cal L}\tau} \approx  e^{{\cal L}_2\tau/2} e^{{\cal L}_1\tau} e^{{\cal L}_2\tau/2}
\end{equation}
%\begin{equation}
%{\cal L} = {\cal L}_{1} + {\cal L}_{2} + {\cal L}_{3},
%\label{liouOpSplit}
%\end{equation}
%with
%\begin{equation}
%{\cal L}_{1} = {\cal L}_{hs} + \sum_{i}\frac{p_{i}}{m_{i}}\frac{\partial}{\partial q_{i}}
%-\sum_{i}\frac{\partial}{\partial q_{i}}V_1(q)
%\frac{\partial}{\partial p_{i}},
%\label{liouOpiLfirst}
%\end{equation}
%%\begin{equation}
%{\cal L}_{2} = -\frac{\partial}{\partial q_{i}}V_2(q)\frac{\partial}{\partial p_{i}}
%\label{liouOpiLsnd}
%\end{equation}
%and
%\begin{equation}
%{\cal L}_{3} = - \sum_{i}p_{i}\xi\frac{\partial}{\partial p_{i}}
%+\frac{1}{Q}\left(\sum_{i}\frac{p_{i}}{m_{i}} - gkT\right)
%\frac{\partial}{\partial \xi} + \xi\frac{\partial}{\partial \eta}.
%\label{liouOpiL3}
%\end{equation}
%can be used.
%Using a Trotter factorization gives
%\begin{equation}
%e^{\displaystyle {\cal L}\tau} = e^{\displaystyle {\cal L}_3\tau/2}e^{\displaystyle {\cal L}_2\tau/2}
%e^{\displaystyle {\cal L}_1\tau}e^{\displaystyle {\cal L}_2\tau/2}e^{\displaystyle {\cal L}_3\tau/2} + {\cal O}(\tau^3)\; .
%\label{trotterNH}
%\end {equation}

\section{Collision Verlet with a Nos\'{e}-Poincar\'{e} Thermostat}

The Nos\'{e}-Hoover formulation of constant-temperature dynamics is non-Hamiltonian in structure,
thereby preventing the use of symplectic integration schemes, which, for systems with continuous
potentials, can be shown to enhance long-term
stability\cite{Sanz-Serna95}.  Recently, Bond, Leimkuhler, and Laird\cite{Bond99} have proposed 
a new real-time, but fully Hamiltonian, formulation of the Nos\'{e} constant-temperature dynamics. 
This is accomplished by performing a time transformation, not to the Nos\'e equations of motion as
with Nos\'e-Hoover, but directly to the Hamiltonian using a Poincar\'{e} time 
transformation, as follows:
\begin{equation}
{\cal H}_{NP} = s({\cal H}_{Nos\acute{e}} - {\cal H}_0),
\label{Trans}
\end {equation}
where $H_0$ is the initial value of ${\cal H}_{Nos\acute{e}}$.
Combining equations (\ref{HNose}) and (\ref{Trans}) the Nos\'{e}-Poincar\'{e} 
thermostat Hamiltonian of a physical system consisting of N particles is 
expressed as following
\begin{equation}
{\cal H}_{NP} =s\left ( \sum_{i} \frac{\tilde{p}_{i}^2}{2m_{i}s^2}+
V_c(q) + \frac{\pi^2}{2Q} + gkT\ln s - {\cal H}_0 \right) .
\label{Nose-Poincare}
\end{equation}
In order to sample the correct canonical distribution, the constant $g$ is taken 
to be the number of degrees of freedom\cite{Bond99}, $g=N_f$.
The equations of motion are 
\begin{equation}
\dot{q}_{i} = \frac{\tilde{p}_{i}}{m_{i}s},\; \;  \dot{s} = s\frac{\pi}{Q},
\label{eqMotNosePoint}
\end{equation}
\begin{equation}
\dot{\tilde{p}}_{i} =-s\frac{\partial}{\partial q_{i}}V_{c}(q), \; \;
\dot{\pi} = \sum_{i}\frac{\tilde{p}_{i}^2}{m_{i}s^2} - gkT -\Delta{\cal H},
\label{eqMotNosePointsecond}
\end{equation}
\begin{equation}
\Delta{\cal H} = \sum_{i} \frac{\tilde{p}_{i}^2}{2m_{i}s^2}+ V_c(q) + 
\frac{\pi^2}{2Q} + gkT\ln s - {\cal H}_0. 
\label{eqMotNosePointthird}
\end{equation}
Note that, the exact solution to Nos\'e-Poincar\'e equations of motion generates trajectories that
are identical to that generated by the Nos\'e-Hoover scheme, exactly solved.  It is in 
the construction of approximate numerical methods that these two approaches differ.

For the present case, we write the Nos\'{e}-Poincar\'{e} thermostat pseudo-Hamiltonian 
(see Sect. 2) for a mixed hard-core/continuous potentials system 
\begin{eqnarray}
{\cal H}_{NP} &=&s\left( \sum_{i} \frac{\tilde{p}_{i}^2}{2m_{i}s^2}+
V_{hs}(q) + V_c(q) + \frac{\pi^2}{2Q}\right.\nonumber \\
&& \left. + gkT\ln s - {\cal H}_0\right).
\label{HSNose-Poincare}
\end{eqnarray}
There are a variety of ways in which one can construct numerical integration algorithms
using this Hamiltonian. To this end, we first consider two ways of splitting the
overal NP Hamiltonian::
\begin{description}
\item[Splitting I]

\begin{eqnarray}
{\cal H}_{1} = s \left( \sum_{i} \frac{\tilde{p}_{i}^2}{2m_{i}s^2}+ V_{hs}(q) + V_1(q)\right. \nonumber \\
\left. + gkT\ln s - {\cal H}_0 \right) \\
 {\cal H}_{2} = s \left( V_2(q) + \frac{\pi^2}{2Q}\right) 
\end{eqnarray}

\item[Splitting II]
\begin{eqnarray}
{\cal H}_{1} &=& s \left( \sum_{i} \frac{\tilde{p}_{i}^2}{2m_{i}s^2}+ V_{hs}(q) + V_1(q)
- {\cal H}_0 \right) \\
{\cal H}_{2} &=& s \left( V_2(q) + \frac{\pi^2}{2Q} + gkT\ln s \right)
\end{eqnarray}
\end{description}
A Trotter factorization of the flow map (Eq.~\ref{trotter}) is applied to each splitting.
To approximate the flow map generated by ${\cal H}_1$, we employ the Collision
Verlet Scheme given in Eq.~\ref{H1flowapprox} to integrate the system from collision to
collision under the influence of the short-range potential. Since $s$ is a constant in the
dynamics generated by ${\cal H}_1$ in both splittings, the St\"ormer-Verlet
algorithm can be used to integrate the trajectory between collisions, with the collision time
being calculated as described in the Appendix. For splitting I, St\"ormer-Verlet gives
\begin{eqnarray}
\tilde{p}_{i,n+1/2}& =& \tilde{p}_{i,n+1/2} - \frac{\tau}{2}
s_n\frac{\partial}{\partial q_{i}}V_1(q_n) \\
\pi_{n+1/2} &=& \pi_{n+1/2} + \frac{\tau}{2}
\bigg [\sum_i\frac{1}{m_i}\left(\frac{\tilde{p}_{i,n+1/2}}{s_n}
\right)^2  \nonumber\\
&&- \Delta H\left(q_{n},\tilde{p}_{i,n+1/2}, s_n\right)\bigg ] \\
q_{i,n+1}& =& q_{i,n} +  \tau\frac{\tilde{p}_{i,n+1/2}}{m_is_n} \\
\pi_{n+1} &=& \pi_{n+1/2} +  \frac{\tau}{2}\bigg[\sum_i \frac{1}{m_i}
\left(\frac{\tilde{p}_{i,n+1/2}}{s_n}\right)^2  \nonumber\\
&&- \Delta H\left(q_{n+1},\tilde{p}_{i,n+1/2},s_n
\right)\bigg] \\ \label{H1psupdate}
\tilde{p}_{i,n+1} &=& \tilde{p}_{i,n+1/2} - \frac{\tau}{2} s_n\frac{\partial}
{\partial q_{i}}V_1(q_{n+1}) \;.
\end {eqnarray}
The equations for Splitting II can be similarly generated. 

In both Splittings I and II the integration of ${\cal H}_2$ is complicated by the presence of
both $s$ and its conjugate momentum $\pi$, but here we consider two possible approaches: 
\begin{description}
\item[\underline{${\cal H}_2$ Integration Method 1}:]
Since the Hamiltonian here is non-separable, the Generalized \linebreak[4] Leapfrog \cite{Hairer94,Sun93,Bond99} scheme,
a fully symplectic extension of the St\"ormer-Verlet algorithm for non-seperable Hamiltonians, can be used. 
The integration for Splitting I for timestep $\tau$ is
\begin{equation}
\tilde{p}_{i,n+1/2} =  \tilde{p}_{i,n} - \frac{\tau}{2}s_n
\frac{\partial}{\partial q_{i}}V_2(q_n)
\end {equation}
\begin{equation}
\pi_{n+1/2} = \pi_{s,n} - \frac{\tau}{2}\left(gkT +
\Delta H_2\left(q_n,s_n,\pi_{n+1/2}\right)\right)
\label{GLAquadratic}
\end {equation}
\begin{equation}
s_{n+1} = s_n + \frac{\tau}{2}\left(s_n + s_{n+1}\right)
\frac{\pi_{n+1/2}}{Q},
\end {equation}
\begin{equation}
\pi_{n+1} = \pi_{n+1/2} - \frac{\tau}{2}\left(gkT +
\Delta H_2 \left(q_n,s_{n+1},\pi_{n+1/2}\right)\right)
\end{equation}
\begin{equation}
\tilde{p}_{i,n+1} = \tilde{p}_{i,n+1/2} -\frac{\tau}{2}s_{n+1}
\frac{\partial}{\partial q_{i}}V_2(q_n)
\end {equation}
The above integration is explicit. Eq.~\ref{GLAquadratic} requires
the solution of a scalar quadratic equation for $\pi_{n+1/2}$. Details of how
to solve this equation without involving subtractive cancellation can be 
found in Ref.~\cite{Bond99}. The application of Method 1 for the ${\cal H}_2$
in Splitting II is similar and straightforward.

\item[\underline{${\cal H}_2$ Integration Method 2}:]
Instead of using Generalized Leapfrog, we employ a splitting of ${\cal H}_2$
\begin{equation}
{\cal H}_2 = {\cal H}_2^{(1)} + {\cal H}_2^{(2)}.
\end {equation}
For Splitting I, we use 
\begin{eqnarray}
{\cal H}_2^{(1)}& =& \frac{s\pi^2}{2Q}  \\
{\cal H}_2^{(2)} &=& sV_2(q) \;.
\end{eqnarray}
Since no conjugate pair appears in ${\cal H}_2^{(2)}$, its dynamics for a 
timestep $\tau$ is straightforward
\begin{eqnarray}
\tilde{p}_{i,n+1} &=& \tilde{p}_{i,n} -\tau s_{n}\frac{\partial}{\partial q_{i}}V_2(q_n)\\
\pi_{n+1} & =&  \pi_{n} -\tau V_2(q_n)
\end{eqnarray}
Only equations involving variables $p$ and $\pi$ are shown above because $q$ 
and $s$ are constants of motion.

The solution of the dynamics of ${\cal H}_2^{(1)}$ involves a conjugate pair
$s$ and $\pi$, but it can be solved exactly~\cite{Nose01}. Thus the time 
evolution of ${\cal H}_2^{(1)}$ for the timestep $\tau$ is
\begin{equation}
s_{n+1} = s_n\left(1 + \frac{\pi_{n}}{2Q}\tau\right)^2
\label{splitNosefirsteq}
\end {equation}
\begin{equation}
\pi_{n+1} = \frac{\pi_{n}}{1 + \frac{\pi_{n}}{2Q}\tau}.
\label{splitNosesecondeq}
\end {equation}
Here, it is $q$, and $\tilde{p}$ that are constants of motion.
Again, the application of Method  2 for  Splitting  II is similar and straightforward.
\end {description}
Combining the two overall splittings for the NP Hamiltonian with the two methods for
integrating ${\cal H}_2$, gives a total of 4 proposed algorithms for the Nos\'e-Poincar\'e
Collision-Verlet (NPCV) method. These are
\begin{itemize}
\item {\bf NPCV1:}  Splitting I + ${\cal H}_2$ integration method 1
\item {\bf NPCV2:}  Splitting I +${\cal H}_2$ integration method 2
\item {\bf NPCV3:}  Splitting II +${\cal H}_2$ integration method 1
\item {\bf NPCV4:}  Splitting II +${\cal H}_2$ integration method 2
\end{itemize}
In the next section we test these four algorithms for a model system and compare them with
each other and with the Nos\'e-Hoover Collision Verlet (NHCV) method outlined in the 
previous section.
\section{Numerical Experiments on a Model Potential}

We test the various algorithms for NVT Collision Verlet proposed in this paper using a system
of hard-spheres with an attractive inverse-sixth-power 
continuous potential,
\begin{equation}
v_{c} = -\epsilon \left (\frac{\sigma}{q} \right )^6 \; ,
\end{equation}
where $\sigma$ is the hard-sphere diameter.
The potential is truncated at the distance $q_c = 2.5\sigma$ and, to ensure
its continuity, it is shifted and smoothed 
so that potential and the force vanish beyond the cutoff distance. 
We split the above potential into short and long-range parts, as prescribed in 
Ref.\cite{Houndonougbo00}, with $q_1$ and $q_2$ as input parameters. 
\begin{figure}
\centerline{\resizebox{9cm}{9cm}{\includegraphics*[0cm,0cm][16cm,14cm]{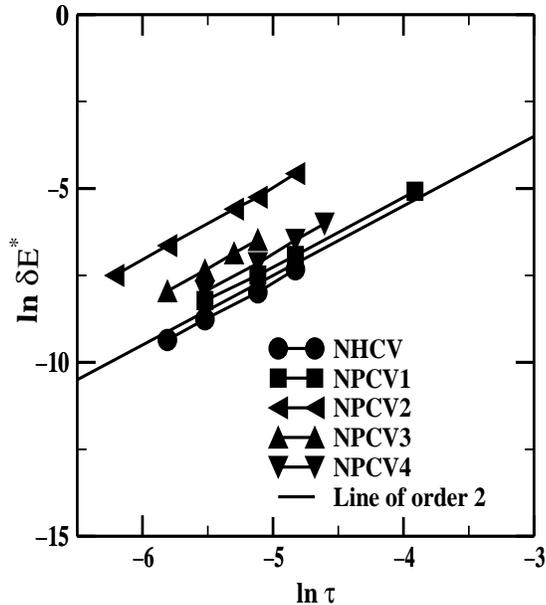}}}
\caption{order of accuracy of the NHCV algorithm and  NPCV algorithms 1 to 4.
Comparison is made with a line of order 2.}
\label{fig1}
\end{figure}
\begin{figure}
%\centerline{\includegraphics[0cm,-5cm][20cm,20cm]{fig3.ps}}
\centerline{\resizebox{10cm}{10cm}{\includegraphics*[0cm,0cm][20cm,20cm]{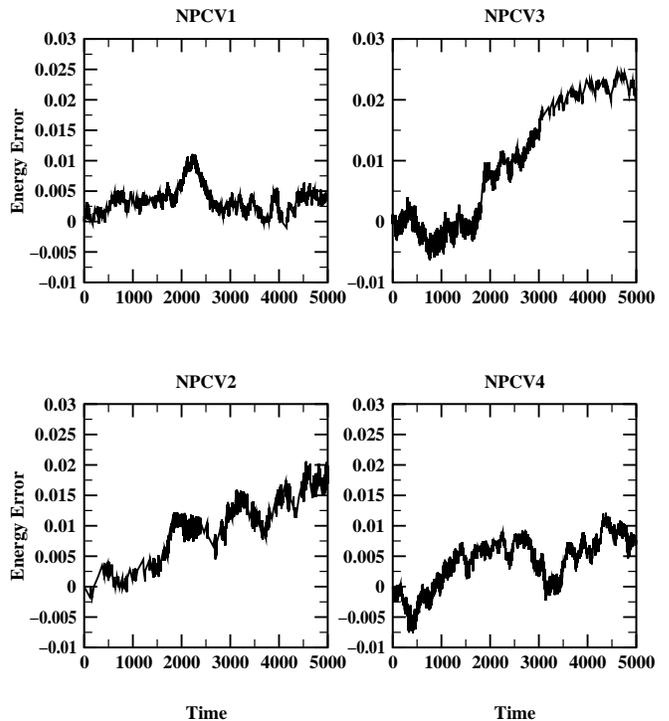}}}
\caption{Energy conservation in a long simulation run ($10^6$ time steps) for NPCV algorithms 1 to 4. }
\label{fig2}
\end{figure}
\begin{figure}
%\centerline{\epsfig{file= fig1.ps , bbllx=-0cm,bblly=0cm,bburx=18cm,bbury=15cm}}
\centerline{\resizebox{9cm}{9cm}{\includegraphics*[0cm,0cm][18cm,18cm]{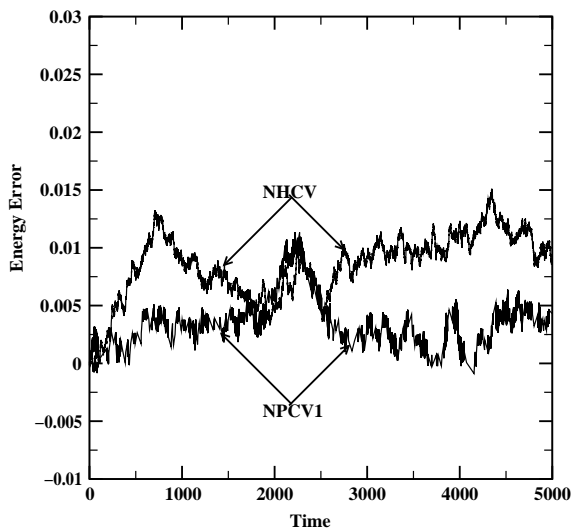}}}
\caption{Energy versus time in a long simulation run ($10^6$) using the NHCV and NPCV1 algorithms}
\label{fig3}
\end{figure}
\begin{figure}
%\centerline{\includegraphics[0cm,-5cm][20cm,20cm]{fig4.ps}}
\centerline{\resizebox{11cm}{11cm}{\includegraphics*[0cm,-5cm][20cm,20cm]{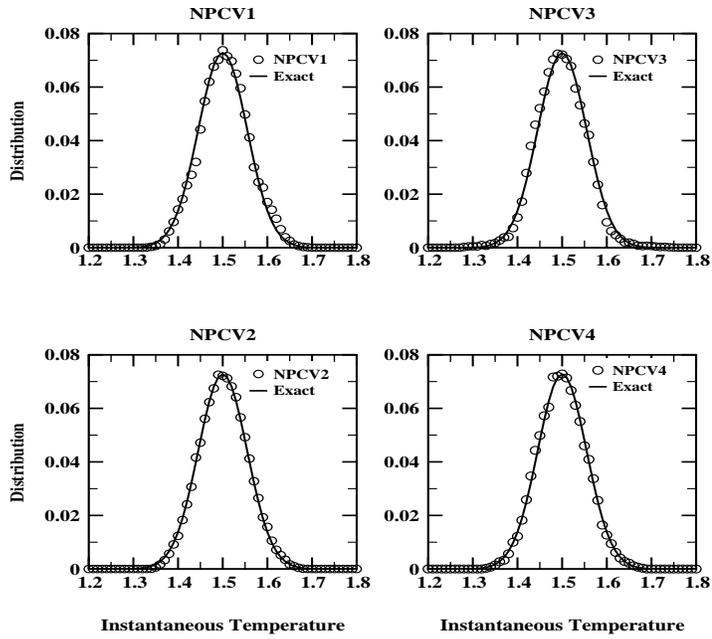}}}
\caption{Instantaneous temperature distributions for the NPCV algorithms 1 to 4.
In each, the exact canonical distribution is shown as a solid line.}
\label{fig4}
\end{figure}
\begin{figure}
%\centerline{\epsfig{file= fig2.ps ,bbllx=3cm,bblly=0cm,bburx=18cm,bbury=9cm}}
\centerline{\resizebox{9cm}{9cm}{\includegraphics*[0cm,0cm][18cm,18cm]{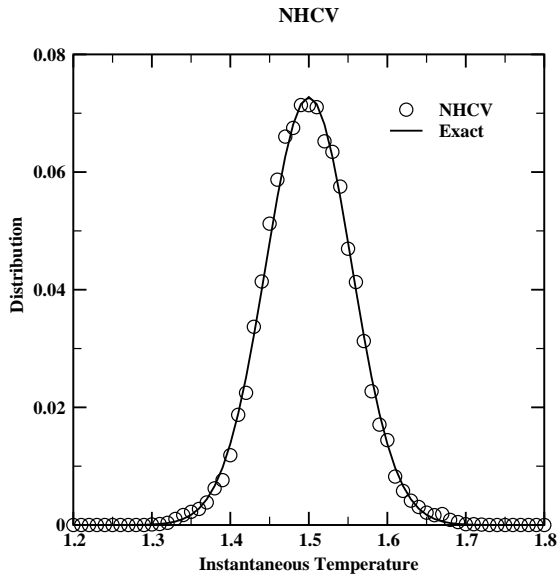}}}
\caption{Instantaneous temperature distribution for the NHCV simulations (circles). The exact canonical distribution is shown as a solid line.}
\label{fig5}
\end{figure}

The MD simulations were carried out on systems of $N = 500$ particles. 
A system of reduced units was chosen so that all quantities are 
dimensionless: as units of distance and energy we used the potential 
parameters $\sigma$ and $\epsilon$, respectively, and the mass of one atom 
as the unit mass. The unit of time is $(m \sigma^2/\epsilon)^{1/2}$. 
An asterisk superscript indicates reduced units. In all simulations 
the density was  $\rho^{\ast} = \rho \sigma^3 = 0.7$ with reduced temperature 
$T^{\ast} = kT/\epsilon = 1.5$. In addition, a cubic box with periodic 
boundary conditions was used. 
In improve efficiency, neighbor (Verlet) 
lists\cite{Allen87} were used  for the evalution of the short range force, the long range force, 
and the collision times. In all of our simulations, we set $g=N_f$ with
$N_f = 3(N-1)$ to correct for the fact that in a molecular-dynamics
simulation the total linear momentum is conserved\cite{Cagin88}.
Each run has was started form an initial configuration produced
after an equilibration run of  200,000 time steps (with $\tau^{\ast} = 0.001$) starting 
from an fcc (face-centered-cube) lattice with the particle velocities chosen from 
a Boltzmann distributuion at $T^{\ast} = 1.5$. 
The initial values of the extended variables in all of the numerical experiments
are set to be $s_0=1$ and $p_{s,0} = 0$ in the case of the Nos\'{e}-Poincar\'{e}
thermostat methods.  In the case of the Nos\'{e}-Hoover method, the initial values of 
the extended variables  are thus $\eta_0 = 0$ and $\xi_0 = 0$.  
%In all runs a
%reduced time step of $\tau^{\ast} = 0.005$ was used.
%and the runs consisted of
%200,000 steps of equilibration and 1,000,000 steps for averages, which would
%correspond to just over a nanosecond of total run time, assuming that a single time step
%corresponds to approximately 1 femtosecond. 

In order to compare the short time accuracy of the methods and verify that each one 
exhibits second-order global error, we show in Figure 1 a log-log plot of the 
maximum energy error for a run of total length $t^{\ast} = 12$ for each method as a 
function of time step, $\tau$. For comparison, a line of slope 2 is plotted to show
that the global error for each method is second order, as required. In these runs the
thermostat mass $Q$ was set to 1.0. Note that, due to
the discontinuous nature of the dynamics, the second order global error is not simply
a consequence of the time-reversibility of the algorithms, but it also a direct result of
the particular potential splitting we have chosen\cite{Houndonougbo00}. From Figure 1 we
see that for short runs, the Nos\'e-Hoover based method has the smallest error
constant. 

For molecular-dynamics simulation the stability during long runs is more important that
the order or short-term behavior of the algorithm. To test these we plot the energy
trajectory, $\delta E = E(t) - E(t=0)$, versus time for each of our methods 
using $10^{6}$ time steps of length 
$\tau^{\ast} = 5\times 10^{-3}$ (total time 5000). Figure 2 shows this plot for each
of the 4 Nos\'e-Poincar\'e based methods discussed in the previous section. For this system,
NPCV methods 2 and 3 exhibit significant drift whereas methods 1 and 4 are more stable
for long time trajectories. The same plot for the Nos\'e-Hoover method presented in section
3 is shown in Figure 3 with the plot for NPCV method 1 shown for comparison. The NPCV method
1 has slightly better energy conservation for this system than the Nos\'e-Hoover Collision
Verlet algorithm, which is comparable to NPCV method 4, but the differences are small 
and could change depending on the system. 

The algorithms presented here are designed to give a canonical distribution of phase
space points. A useful check of this is to examine the distribution of instantaneous
temperature (as defined for a system with zero total momentum)
\begin{equation}
\hat{T} = \frac{2}{3(N-1)}\sum_i^N \frac{p^2_i}{2 m} 
\end{equation}
A canonical distribution in momenta requires that this quatitiy be Gaussian distributed
about the target temperature $T$ with a variance of $\frac{2T^2}{3(N-1)}$. In Figure 4 is
plotted the temperature distributions for the 4 NPCV algorithms using a thermostat
mass of 10 measured during runs of 270,000 time steps ($\tau^{\ast} = 5\times 10^{-3}$) after
equilibration.  Figure 5 shows the same quantity for the Nos\'e-Hoover Collision Verlet
method. Comparison with the theoretical distribution, shown as a solid line in
each plot, indicates that the canonical distribution is well reproduced by all proposed 
algorithms. 

\section{Conclusion}

In this work we have developed several algorithms, based on the extended Hamiltonian
thermostat of Nos\'e, to perform constant temperature ($NVT$) molecular-dynamics simulations
of systems with mixed hard-core/continuous potentials. The methods are extentions of
our recently developed Collision Verlet method\cite{Houndonougbo00} for constant energy ($NVE$) 
MD simulation of such systems. These new methods, to our knowledge, represent the first
viable canonical molecular-dynamics simulation methods for hybrid discontinous/continuous
potentials. 

Specifically, five new algorithms have been presented and tested. The first algorithm,
the Nos\'e-Hoover Collision Verlet (NHCV) algorithm, is based on application of the 
Nos\'e-Hoover thermostat\cite{Hoover85} to the Collision Verlet scheme. The other 4 algorithms 
presented are based on the Nos\'e-Poincare formulation of real-time Nos\'e dynamics. 
These Nos\'e-Poincar\'e Collision Verlet methods differ from one another in the details of
the numerical scheme used to  integrate the equations of motion. All methods were shown
to give second-order global error in test simulation with the NHCV method having the smallest
error constant for short-time simulations. The NHCV algorithm and two of 
the presented NPCV algorithms (NPCV1 and NPCV4) were found to exhibit good stability in
long time simulations involving 500 hard-sphere particles with attractive inverse-sixth-power
tails. In addition, all methods were shown to correctly reproduce the canonical distribution 
of instantaneous temperature (kinetic energy). Note that, if the continuous potential is
set to zero, the presented methods also provide a way of performing canonical, as opposed
to isokinetic, hard-sphere molecular-dynamics simulations. 
\vskip 0.5cm
\noindent 
\begin{acknowledgements}
The authors wish to thank Professor Benedict Leimkuhler for helpful discussions 
and gratefully acknolwedge the National Science Foundation for financial support under grant CHE-9970903.
In addition, we thank the Kansas Center for Advanced Scientific Computing for use of
their computational facilities.
\end{acknowledgements}
\noindent
\appendix*
\section{Calculation of time to next collision}

In this appendix we address the issue of the collision time calculation for
mixed hard-core/continuous potentials systems. The quartic equation for the
collision condition (Eq.~\ref{coll_cond}), is solved for all pairs of particles and the smallest
positive root is located as the time to the next collision.
For mixed hard-core/continuous potentials systems, this is time-consuming
operations since collision times for all pairs must be recalculated after
each collision. In addition,  Eq.~\ref{coll_cond} is quartic and difficult
to solve. As we said in section 2, the quartic equation must be solved
accurately to give the nearest root to zero in order to make sure that no
collisions are missed.

In ref.\cite{Houndonougbo00}, we employed  Laguerre's method\cite{NumRes} for collision
time calculation for mixed hard-core/continuous potentials systems. The
method is sufficient for all but the very smallest timesteps studied. But
the method turns out to be very slow. This because for any given time interval
and pair of particles, all the four complex roots need to be calculated.
Also Laguerre's method deals with complex arithmetic. In this appendix, we
propose a time saving collision time calculation method for collision verlet.
This method is based on a Cauchy indices of a Sturm sequence\cite{Henrici74} of a 
real polynomial in a real interval.

The Cauchy index is an integer that can be associated with any real rational
function and any interval whose end points are not the function poles.
Let $r$ be a rational function. The {\bf Cauchy index}, $I^{\beta}_{\alpha}r(x)$,
of $r$  for the interval $[\alpha, \beta]$  is by definition the number of
jumps of the function r from $+\infty$ to $-\infty$ on the interval
$[\alpha, \beta] $.
The Cauchy index can be calculated for any real polynomial that forms a
{\bf Sturm sequence}, $\{ f_0,f_1,...,f_m\}$, for
the interval $[\alpha, \beta]$. The definition of the Sturm sequence of a
real polynomials can be found in ref.\cite{Henrici74}. The connection between
the Cauchy index and the number of sign changes, $v(x)$ for arbitrary real $x$,
 in the numerical sequence ,$\{ f_0,f_1,...,f_m\}$, is given by
the following result due to Sturm\cite{Sturm1835}.

-----------------------------------------------------------------------
\newtheorem{theorem}{Theorem}
\begin{theorem}
Let the real polynomials ,$\{f_0,f_1,...,f_m\}$ form a Sturm
sequence for the interval $[\alpha, \beta]$, $\alpha \le \beta$. Then
\begin{equation}
I^{\beta}_{\alpha}\frac{f_1}{f_0} = v(\alpha)-v(\beta).
\end{equation}
\label{sturm}
\end{theorem}
-----------------------------------------------------------------------

Using this theorem we can write the number of real roots for a given polynomial 
$p$ in any real interval $[\alpha, \beta]$ in terms of the Cauchy index
\begin{equation}
I^{\beta}_{\alpha}\frac{p'}{p_0} = v(\alpha)-v(\beta).
\label{cauchyindex}
\end{equation}
of the sequence $\{p_k\}$, generated by the Euclidean algorithm\cite{Henrici74} using the
starting polynomials $p_0:=p, p_1:=p'$, with
$p'$ being the first derivative of the polynomial $p$.
The elements of the rest of the sequence are linked by the relations
\begin{eqnarray}
p_0(x) = q_1(x)p_1(x)-p_2(x),\\
p_1(x) = q_2(x)p_2(x)-p_3(x),\\
\vdots  \nonumber\\
p_{k-1}(x) = q_k(x)p_k(x)-p_{k+1}(x),\\
\vdots  \nonumber\\
p_{m-1}(x) = q_m(x)p_m(x).
\end{eqnarray}
The Euclidean algorithm also furnishes information about the multiplicity of
the zeros. $x_0$ is a zero of multiplicity $k$ of $p$ if and only if
it is a zero of multiplicity $k-1$ of $p_m$.
We are now able to develop a collision time calculation method for Collision
Verlet.

 From the above, the first step for Collision Verlet collision time calculation is
to determine in a given time interval the number of real roots by calculating
the Cauchy index for the time interval. This means that we need an algorithm
for polynomial division. The main problem with polynomials division is that
the bitlenght of coefficients in the sequence can increase dramatically and
also, because we are dividing, in some cases the denominator can vanish.
To solve this problem, we use the {\bf Sturm-Habicht} pseudodivisions
subresultant (PRS) method\cite{Akritas89}. The members of the polynomial
remainder sequence $p_1(x),p_2(x),p_3(x),...,p_h(x)$
\begin{equation}
{l_c[p_{i+1}(x)]}^{n_i-n_{i+1}+1} p_i(x) = p_{i+1}(x)q_i(x) - \beta_i p_{i+2}(x),
\label{Habicthfirst}
\end{equation}
\begin{equation}
deg[p_{i+2}(x)] \le deg[p_{i+1}(x)]
\end{equation}
where $i=1,2,..,h-1$, for some h, $n_i = deg[p_i(x)]$, and $l_c[p_i(x)]$ is the leading coeficient of $p_i$. The different values
of $\beta_i$ are
\begin{eqnarray}
\beta_1 &=&(-1)^{n_1-n_2+1},\\
\beta_i& =& (-1)^{n_i-n_{i+1}+1}l_c[p_i(x)]
\cdot H^{n_i-n{i+1}}_i, \nonumber\\
&&i = 2,3,...,h-1,\\
\label{betafirst}
H_2& =& \{l_c[p'_2(x)]\}^{n_1-n_2},\\
\label{betasec}
H_i& =& \{l_c[p_i(x)]\}^{n_{i-1}-n_i}H^{1-(n_{i-1}-n_i)}_{i-1}, \nonumber\\
&&i=3,...,h-1
\label{betathird}
\end{eqnarray}
Let
\begin{equation}
p(x) = ax^4 + bx^3 +cx^2 + dx + e,
\label{collieq}
\end{equation}
be the quartic polynomial obtained from the collision condition of eq. (\ref{coll_cond}),
and $\{p_1,p_2,p_3,p_4,p_5\}$ its Sturm-Habitch sequence determined by using
eq. \ref{Habicthfirst}. We now determine the number of real roots of the 
equation $p(t) = 0$ in a given time interval by calculating its
Cauchy index, Eq.~\ref{cauchyindex}. If there is only one root, then we use
Newton-Raphson method~\cite{NumRes}
to approximate the root. If there is more than one root, then we combine
bisection method~ \cite{NumRes} and root counting method to isolate the time
interval containing the smalest root.

This method for solving for the shortest collision time is quite efficient giving a factor of
20 speed-up from our previous simulations using the Laguerre method~\cite{Houndonougbo00}, 
primarily because we no longer calculate all four roots of the quadratic equation
and avoid complex arithmetic. 
\bibliography{master}
\bibliographystyle{laird_notitle}
\end{document}